\documentstyle[12pt]{article}
\begin{document}
\baselineskip 18pt
\def\today{\ifcase\month\or
 January\or February\or March\or April\or May\or June\or
 July\or August\or September\or October\or November\or December\fi
 \space\number\day, \number\year}
\def\thebibliography#1{\section*{References\markboth
 {References}{References}}\list
 {[\arabic{enumi}]}{\settowidth\labelwidth{[#1]}
 \leftmargin\labelwidth
 \advance\leftmargin\labelsep
 \usecounter{enumi}}
 \def\newblock{\hskip .11em plus .33em minus .07em}
 \sloppy
 \sfcode`\.=1000\relax}
\let\endthebibliography=\endlist
\renewcommand{\thefootnote}
{\fnsymbol{footnote}}
\begin{titlepage}
%
   
\hspace*{10.0cm}OCHA-PP-113
\  \
\vskip 0.5 true cm 
\begin{center}

{\large {\bf 
Electroweak Baryogenesis and Neutron Electric Dipole Moment
from Supersymmetry
\footnote{To be published in the Proceedings of the 3rd RESCEU  
International Symposium on Particle Cosmology.}   
}}
\vskip 2.0 true cm

Mayumi AOKI \footnote{Research Fellow of the Japan 
Society for the Promotion of Science.} \\
{\it    Graduate School of Humanities and Sciences \\
Ochanomizu University, 
Bunkyo-ku, Tokyo 112-8610,  JAPAN \\ 
amayumi@fs.cc.ocha.ac.jp}\\
\vskip 0.5 true cm 
Noriyuki OSHIMO\\
{\it Institute for Cosmic Ray Research \\
University of Tokyo, Tanashi, Tokyo 188-0002, JAPAN \\
oshimo@icrr.u-tokyo.ac.jp}\\
\vskip 0.5 true cm 
Akio SUGAMOTO \\
{\it    Department of Physics \\
Ochanomizu University, Bunkyo-ku, Tokyo 112-8610, JAPAN \\
sugamoto@phys.ocha.ac.jp}
\end{center}

\vskip 3.0 true cm

\centerline{\bf Abstract}
\medskip
 
     Baryogenesis at the electroweak phase transition is discussed
within the framework of the supersymmetric standard model.
Implications of baryon asymmetry for the electric dipole moment
of the neutron are also studied.

\medskip

\end{titlepage}

\def\gsim{{\mathop >\limits_\sim}}
\def\lsim{{\mathop <\limits_\sim}}
\def\r2{\sqrt 2}
\def\sw2{\sin^2\theta_W}
\def\v#1{v_#1}
\def\tb{\tan\beta}
\def\c2b{\cos 2\beta}
\def\sq{\tilde q}
\def\st{\tilde t}
\def\m#1{{\tilde m}_#1}
\def\mH{m_H}
\def\mg{{\tilde m}_g}
\def\M{\tilde M}
\def\mgr{m_{3/2}}
\def\dW{\delta_W}
\def\vW2{v_W^2}
\def\muB{\mu_B}

     Astronomical observations indicate that there exist more baryons 
than anti-baryons in our universe.
The ratio of baryon number to entropy  
is observed at present to be $\rho_B/s = (2-9)\times 10^{-11}$.
This asymmetry may have been generated at the electroweak phase
transition of the universe,  
since all the necessary conditions for baryogenesis seem
to be satisfied within the framework of electroweak theories.
The standard model (SM), however, cannot account for 
the asymmetry quantitatively.
In this note, we report the possibility of baryogenesis 
by the charge transport mechanism in 
the supersymmetric standard model (SSM) based on
$N$=1 supergravity. 

     The SSM has several complex parameters in addition to the
Yukawa coupling constants for quarks.
We can take for new physical complex parameters,
without loss of generality,
a Higgsino mass parameter $\mH$ for the bilinear term
of the Higgs superfields in the
superpotential and a dimensionless coupling
constant $A$ for the trilinear terms of 
scalar fields which break supersymmetry softly.
We express these parameters as
$\mH = |\mH|\exp(i\theta)$ and $A = |A|\exp(i\alpha)$.  
The complex phases $\theta$ and $\alpha$ become 
new origins of $CP$ violation. 

     The new sources of $CP$ violation in the SSM give
contributions to the electric dipole moment (EDM) of 
the neutron at the one-loop level 
through diagrams in which charginos, neutralinos or gluinos 
propagate together with squarks.
The experimental upper bound of the neutron EDM is approximately
$1 \times 10^{-25}e$cm.
1) If the $CP$-violating phase $\theta$ is of order unity, the neutron  
EDM receives a contribution dominantly from the chargino diagram.
From the experimental constraint, the masses of squarks are predicted
to be larger than 1 TeV, while the chargino masses can be of order 100 GeV.
In this case, charginos can mediate the charge transport mechanism.
2) If the phase $\theta$ is sufficiently smaller than unity,
the chargino contribution to the EDM is negligible.
The neutron EDM receives dominant contributions
from the gluino and neutralino diagrams,
leading to relaxed constraints on the masses of $R$-odd particles. 
The squarks are allowed to have masses of order 100 GeV,
even if the phase $\alpha$ is of order unity. 
Then, $t$ squarks assume the role of the mediators for the
charge transport mechanism.

     At the electroweak phase transition, if it is first order,
bubbles of the broken phase nucleate in the symmetric phase.
The charged gauginos $\lambda$ and the charged Higgsinos $\psi$ are in
mass eigenstates in the symmetric phase.
In the broken phase and the bubble wall, they are mixed to 
form charginos as mass eigenstates, owing to non-vanishing 
vacuum expectation values (vevs) $\v1$ and $\v2$ of Higgs bosons.
At the bubble wall, $\lambda$ coming from the symmetric phase
can be reflected to become $\psi$, and vice versa.
The charginos in the broken phase can be transmitted to 
the symmetric phase and become $\lambda$ or $\psi$.
In these processes, 
$CP$ violation is induced by the phase $\theta$ contained in
the chargino mass matrix, making differences in reflection
and transmission probabilities between $CP$-conjugate reactions.
A net flux of hypercharge is then emitted into the symmetric phase.
In the symmetric phase, the chemical potential of baryon number is related to
a hypercharge density through equilibrium conditions.
The induced hypercharge density in front of the wall makes 
the chemical potential of baryon number non-vanishing,
which becomes a bias favoring a non-vanishing value of the baryon number.
Baryogenesis then occurs, since the baryon number can change
through an electroweak anomaly at a non-negligible rate,
$\Gamma=3 \kappa (\alpha_WT)^4$, where $T$ denotes   
a temperature at the phase transition and $\kappa \sim 0.1-1$.
We take $\kappa=1$ for numerical evaluations.

     We calculated the baryon asymmetry for $\theta \sim 1$ and
the chargino masses of order 100 GeV.
The obtained ratio is $\rho_B/s \sim 10^{-10}- 10^{-11}$,
which is compatible with the observed value.
The ratio $\v2/\v1$ does not affect much the result.  
The velocity and width of the bubble wall are taken to be 0.1$-$1 and
$1/T-5/T$, respectively.
If the phase $\theta$ becomes of order 0.1, 
the resultant ratio is at most of order $10^{-11}$.
In the parameter ranges where a sufficient amount of 
the baryon asymmetry is produced, 
the neutron EDM is consistent with its experimental constraint,
if the squark masses $M_{\tilde q}$ are larger than 1 TeV.
For 1 TeV $\lsim M_{\tilde q} \lsim$ 10 TeV, 
the neutron EDM is predicted to be $10^{-25}-10^{-26}e$cm.

     The baryon asymmetry can also be induced by $t$ squarks.  
At the electroweak phase transition, the left-handed $t$ squark
$\st_L$ and the right-handed $t$ squark $\st_R$ are in
mass eigenstates in the symmetric phase,
while non-vanishing vevs cause mixings of $\st_L$ and $\st_R$
in the bubble wall and the broken phase.
At the bubble wall, $\st_L$ coming from the symmetric phase
can be reflected to become $\st_R$, and vice versa.
The mass eigenstates of $t$ squarks in the broken phase
can be transmitted to the symmetric phase and become $\st_L$ or $\st_R$.
In these processes, the complex phase $\alpha$ in the mass squared-matrix
of top squarks induces $CP$ violation, causing differences in
reflection and transmission rates between $CP$-conjugate reactions.
A net hypercharge flux is emitted into the symmetric phase.
Then, similarly to the baryogenesis by charginos, the baryon 
asymmetry is produced.  

     We calculated the baryon asymmetry for $\theta \ll 1$, 
$\alpha \sim 1$, and the squark masses of order 100 GeV.
The obtained ratio can be as large as
$5\times 10^{-11}$ in the same ranges for the velocity and width
of the wall as those for the chargino case. 
The ratio $\v2/\v1$ must vary in the wall for $CP$ violation to occur.
If the phase $\alpha$ becomes of order 0.1, 
it is difficult to produce a sufficient degree of asymmetry.
The magnitude of the neutron EDM is below the experimental 
upper bound if 500 GeV $\lsim \m2$, $\m2$ being the SU(2) gaugino mass.
For 500 GeV $\lsim \m2 \lsim$ 1 TeV,
the neutron EDM is predicted to be $10^{-25}-10^{-26} e$cm. 

     We have studied the possibility of baryogenesis in the SSM,
assuming that the electroweak phase transition is strongly first order.
The charge transport mechanism mediated by charginos or 
$t$ squarks can generate the ratio of baryon number to entropy
consistent with its observed value,
provided that the relevant $CP$-violating phase is not
much suppressed and the relevant particles have masses of order 100 GeV.
Then it is likely that the neutron EDM becomes $10^{-25}-10^{-26}e$cm.
If the baryon asymmetry originates in the new source of $CP$
violation in the SSM, the neutron EDM has a magnitude
which can be explored in near-future experiments.

\section*{References}
\noindent
1.	Aoki, M., Oshimo, N. and Sugamoto, A., 1997,
                {\it Prog.\ Theor.\ Phys.} {\bf 98}, 1179. \\
2.	Aoki, M., Sugamoto, A. and Oshimo, N., 1997,
                {\it Prog.\ Theor.\ Phys.} {\bf 98}, 1325.

\end{document}